\documentclass[twocolumn]{aastex7}

\usepackage{CJK}
\usepackage{enumitem}

\begin{document}
\begin{CJK*}{UTF8}{gbsn}

\title{A Systematic Search for Gaseous Debris Disks in DESI Early Data Release White Dwarfs}

\correspondingauthor{Xiaoxia Zhang}
\email{zhangxx@xmu.edu.cn}

\author{Ziying Ma (马紫莹)}
\affiliation{Department of Astronomy, Xiamen University, Xiamen, Fujian 361005, People's Republic of China}
\email{}
\author[orcid=0000-0003-4832-9422]{Xiaoxia Zhang (张小霞)}
\affiliation{Department of Astronomy, Xiamen University, Xiamen, Fujian 361005, People's Republic of China}
\email{zhangxx@xmu.edu.cn}  
\author[0000-0002-2853-3808]{Taotao Fang (方陶陶)}
\affiliation{Department of Astronomy, Xiamen University,  Xiamen, Fujian 361005, People's Republic of China}
\email{}
\author[0000-0003-4874-0369]{Junfeng Wang (王俊峰)}
\affiliation{Department of Astronomy, Xiamen University,  Xiamen, Fujian 361005, People's Republic of China}
\email{} 
\author[0000-0002-8321-1676]{Jincheng Guo (郭金承)}
\affiliation{Department of Scientific research, Beijing Planetarium, Xizhimenwai Road, Beijing 100044, People's Republic of China}
\email{}
\author[0000-0002-3878-5590]{Xiaochuan Jiang (姜小川)}
\affiliation{School of Information Engineering, Fujian Business University, Fujian 350506, People's Republic of China}
\email{}
\author[0000-0002-2419-6875]{Zhi-Xiang Zhang (张志翔)}
\affiliation{College of Physics and Information Engineering, Quanzhou Normal University, Quanzhou, Fujian 362000, People's Republic of China}
\email[]{}
\author[0000-0002-6684-3997]{Hu Zou(邹虎)}
\affiliation{National Astronomical Observatories, Chinese Academy of Sciences, A20 Datun Road, Chaoyang District, Beijing 100012, People's Republic of China}
\email{}

\begin{abstract}

Detecting gaseous debris disks around white dwarfs offers a unique window into the ultimate fate of planetary systems and the composition of accreted planetary material. Here we present a systematic search for such disks through the \ion{Ca}{2} infrared triplet using the Dark Energy Spectroscopic Instrument (DESI) Early Data Release.  
From a parent sample of 2706 spectroscopically confirmed white dwarfs, we identify 22 candidate systems showing tentative emission-line features, which corresponds to a raw occurrence rate of 0.81\%, more than ten times higher than previous estimates. The detected emission lines are predominantly weak and require confirmation by follow-up observations. 
Three of these candidates also exhibit infrared excess in WISE photometry, suggesting a possible coexistence of gas and dust. However, the high candidate rate indicates that most are likely false positives due to telluric residuals or unresolved binaries. 
This work demonstrates the potential of DESI spectra for blind searches of rare circumstellar phenomena. The recently released DESI DR1, with its substantially larger spectroscopic sample, will enable searches for more gaseous disks and provide better insights into their occurrence and nature. 

\end{abstract}


\section{Introduction} 

White dwarfs represent the end states of the vast majority of stars in the Galaxy, including those known to host planetary systems. Their strong surface gravity causes heavy elements to rapidly settle out of the atmosphere on timescales much shorter than their cooling ages, resulting in atmosphere dominated primarily by hydrogen or helium \citep[e.g.,][]{1979ApJ...231..826F, 2009A&A...498..517K}. Nonetheless, a significant fraction (20\%--50\%) of white dwarfs exhibit  photospheric metal pollution, indicating ongoing accretion of circumstellar material \citep[e.g.,][]{2003ApJ...596..477Z, 2010ApJ...722..725Z, 2014A&A...566A..34K}. This material is widely interpreted as debris from planetary systems that  survived post-main-sequence evolution, offering a unique opportunity to study the composition and evolution of extrasolar planetary material \citep[e.g.,][]{2007ApJ...671..872Z, 2012MNRAS.424..333G, 2014AREPS..42...45J, 2016RSOS....350571V}.

Circumstellar debris disks around white dwarfs are believed to form from the tidal disruption of asteroidal or planetary bodies that venture too close to the star, potentially scattered from regions analogous to the asteroid belt or Kuiper Belt in our solar system \citep[e.g.,][]{2003ApJ...584L..91J, 2011MNRAS.414..930B, 2012ApJ...747..148D}. These debris systems are observed in two primary forms. The first is dusty debris disks, detected as infrared excess over the white dwarf's photospheric emission due to thermal radiation from dust grains. Such disks were first identified with the Spitzer Space Telescope and later in larger numbers by the Wide-field Infrared Survey Explorer \citep[WISE; e.g.,][]{2006ApJ...646..474K, 2011ApJS..197...38D, 2014ApJ...786...77B, 2020ApJ...902..127X, 2023ApJ...944...23W}. 
However, due to the relatively large beam size of WISE ($\simeq6\arcsec$), infrared excesses may be contaminated by cooler companions or background sources.  The occurrence rate of dusty debris disks is estimated to be 1\%--4\% among well-characterized white dwarfs samples robustly observed with facilities such as Spitzer \citep[e.g.,][]{2015MNRAS.449..574R, 2019MNRAS.487..133W, 2021ApJ...920..156L}. 
The second and rarer manifestation is gaseous debris disks. These are definitively identified through double-peaked emission lines, particularly the \ion{Ca}{2} infrared triplet ($\lambda\lambda$ 8498, 8542, 8662\,\AA), which is a signature of Keplerian rotation in a gaseous disk \citep[e.g.,][]{2006Sci...314.1908G}. This gaseous component may coexist with dusty debris or occur independently \citep[][]{2012ApJ...750...86B, 2012MNRAS.421.1635F}, and plays a crucial role in understanding the accretion processes and the dynamics of disrupted planetary material \citep[e.g.,][]{2010ApJ...722.1078M, 2016MNRAS.455.4467M}.

Despite the importance of gaseous debris disks, their origin remains debated, with proposed formation mechanisms including sublimation of dusty material and collisions between planetesimals \citep[e.g.,][]{2011ApJ...732L...3R, 2022MNRAS.509.2404B}. 
Since the first detection of such a system, considerable effort has been devoted to identifying the \ion{Ca}{2} triplet emission through various surveys and follow-up studies \citep[e.g.,][]{2007MNRAS.380L..35G, 2008MNRAS.391L.103G, 2012MNRAS.421.1635F, 2012ApJ...751L...4M, 2014MNRAS.445.1878W, 2015ApJ...810L..17G, 2017ApJ...836...71L, 2019Sci...364...66M, 2020ApJ...905....5D, 2020ApJ...905...56M, 2021MNRAS.504.2707G, 2025PASP..137g4202B, 2025MNRAS.537L..72R}.  
However, to date, only about two dozen have been confirmed, corresponding to an occurrence rate of approximately 0.067\% among white dwarfs \citep[e.g.,][]{2020MNRAS.493.2127M}.  
Expanding this sample is essential to determine the occurrence rate, physical properties, and evolutionary pathways of these gaseous disks.
Furthermore, discoveries of gaseous debris disks have largely been serendipitous. Many have been identified incidentally in large-scale spectroscopic surveys such as the Sloan Digital Sky Survey \citep[SDSS; e.g.,][]{2015MNRAS.446.4078K}, which introduces selection biases since white dwarf candidates were often targeted as quasar candidates or blue-excess sources. Other detections have resulted from targeted follow-up spectroscopy of white dwarfs preselected based on infrared excess or atmospheric metal pollution, a process biased towards the brightest systems.

The Dark Energy Spectroscopic Instrument (DESI) survey is conducting an extensive five-year spectroscopic campaign \citep[][]{2016arXiv161100036D}. While its primary goal is cosmological, DESI also obtains spectra of a vast number of stellar objects including white dwarfs, across approximately one-third of the sky. 
The recent Early Data Release (EDR) from DESI has provided a catalog of about $2700$ spectroscopically confirmed white dwarfs, nearly 60\% of which are newly identified \citep{2024MNRAS.535..254M}. This well-defined sample offers a unique opportunity to perform an unbiased census of rare spectral features in white dwarfs, particularly emission lines indicative of circumstellar processes.
In this paper, we perform a systematic search for the \ion{Ca}{2} triplet emission within the DESI EDR white dwarf sample that are only subject to magnitude limit. We report a selection of new emission-line candidates identified through this blind search. For candidates with archival WISE photometry, we further analyze their spectral energy distributions (SEDs) to identify potential infrared excess, a key indicator of circumstellar dust.

The structure of this paper is as follows. In Section~\ref{sec:data}, we describe the DESI EDR white dwarf sample, our methodology for identifying \ion{Ca}{2} emission lines, and our analysis of archival photometry. Section~\ref{sec:results} presents the properties of the resulting candidate sample. Finally, we discuss the implications of our findings, assess potential sources of contamination, and consider the limitations of our method In Section~\ref{sec:discussion}.

\section{Data and analysis}
\label{sec:data}

\subsection{DESI Spectra}

Our analysis is based on a catalogue of 2706 spectroscopically confirmed white dwarfs from DESI EDR \citep{2024MNRAS.535..254M}. This catalog excludes known binary systems, including white dwarf--main-sequence pairs (WD+MS), double degenerates (WD+WD), and cataclysmic variables (CVs). The spectra were obtained using the DESI instrument on the Mayall 4-m telescope at Kitt Peak National Observatory between December 14, 2020 and June 10, 2021. The data cover a spectral range of 3600--9824\,\AA, with a median full width at half-maximum (FWHM) resolution of $\simeq1.8$\,\AA.

We conducted a systematic search for the \ion{Ca}{2} infrared triplet across the entire sample of white dwarfs. 
The spectra were visually inspected for prominent \ion{Ca}{2} triplet emission; none were found. We therefore proceed to search for weaker emission features through spectral fitting. Objects showing tentative evidence of such emission were designated as candidate gaseous debris disks.

The continuum was modeled using a fifth-order polynomial fitted over the wavelength range 7000--9000\,{\AA} and subsequently subtracted from each spectrum. We then focused on the narrower region of 8450--8700\,{\AA} to search for the \ion{Ca}{2} triplet. This was done by comparing the $\chi^2$ values from two models: (i) a continuum-only model, and (ii) a continuum model plus three Gaussian emission components. 
The Gaussian components were constrained to share a common velocity shift relative to the rest-frame wavelengths of the triplet and were assumed to have identical line widths, reflecting their common physical origin. The lower limit on the line width (standard deviation) was set to 1\,\AA, corresponding to an FWHM of 2.35\,\AA, slightly larger than the instrumental resolution of DESI. 
While the continuum parameters were held fixed, the free parameters in the fit included the common velocity shift, common line width, and the amplitudes of three Gaussian components.

A detection of the \ion{Ca}{2} triplet was defined as a significant improvement in the fit upon inclusion of the Gaussian components, specifically a decrease in $\chi^2$ of at least 15.1 relative to the continuum-only model. This threshold corresponds to a 99\% confidence level ($2.58\sigma$) for five free parameters \citep{1986nras.book.....P}. 
White dwarfs satisfying this criterion were selected as candidate hosts of gaseous debris disks.

\subsection{Photometry and SED fitting}

After selecting white dwarfs with candidate gaseous debris disks based on their optical spectra, we further examine whether these systems exhibit infrared excess using archival observations from WISE. 
Our analysis follows the methodology  in our previous work \citep{2023ApJ...944...23W}, which systematically identified  infrared excess in white dwarfs from the LAMOST DR5 released \citep{2022MNRAS.509.2674G}. 
The difference in the present study is that, once infrared excess is detected, we do not attempt to distinguish its origin, e.g., from a cool companion (such as an M-dwarf or a brown dwarf) or a dusty disk, as our primary aim is to detect  emission-line features from gaseous disks.
We summarize the essential steps of the method below and refer readers to the original publication for full details. 

We compiled optical to infrared photometry for each candidate by cross-matching its DESI coordinates with the VizieR database using a radius of $2\arcsec$. The photometric data were drawn from SDSS DR16 \citep[][]{2020ApJS..249....3A}, Pan-STARRS1 \citep[][]{2016arXiv161205560C}, the Two-Micron All Sky Survey \citep[2MASS;][]{2006AJ....131.1163S}, the UKIDSS Large Area Surveys \citep[][]{2007MNRAS.379.1599L} and WISE \citep[][]{2010AJ....140.1868W}. 
For WISE photometry, we give priority to the deeper CatWISE2020 catalog \citep{2021ApJS..253....8M} to maximize sensitivity to faint sources, falling back to ALLWISE \citep{2014yCat.2328....0C} when necessary.  

Theoretical photospheric emission for each white dwarf was predicted based on its effective temperature ($T_{\rm eff}$) and surface gravity ($\log g$), using established cooling models \citep{1995PASP..107.1047B}. 
Adopting Gaia distances, we converted the predicted absolute magnitudes to radiative fluxes in each photometric system. The observed photometry from SDSS and Pan-STARRS1 were used to fit the cooling model, with a free normalization parameter introduced to account for uncertainties in $\log g$ and distance.
An infrared excess was identified if the flux in any WISE band ($W_1-W_4$) exceeded the model prediction by at least $3 \sigma$ in significance.
For systems showing an infrared excess, we incorporated a disk component into the SED fit, adopting the geometrically flat, optically thick disk model in \citet{2003ApJ...584L..91J}. In this model, the disk is irradiated by the white dwarf and reprocesses the emission in infrared. The outer disk radius was fixed at $R_{\rm out} = 80 R_{\rm WD}$, wherein $R_{\rm WD}$ denotes the white dwarf radius; the inner disk radius ($R_{\rm in}$) and the inclination angle ($i$) were treated as free parameters. The lower limit for $R_{\rm in}$ was set by assuming a dust sublimation temperature of 3000\,K. 

To minimize contamination from binary companions or background sources, we inspected Pan-STARRS1 $z$-band images for all candidates with infrared excess and excluded those having a visible companion within $6\arcsec$, the typical beam size of WISE.


\section{Results}
\label{sec:results}

\subsection{Candidate Gaseous Debris Disks}

\begin{figure*}[tbp]
\centering
\includegraphics[width=1\textwidth]{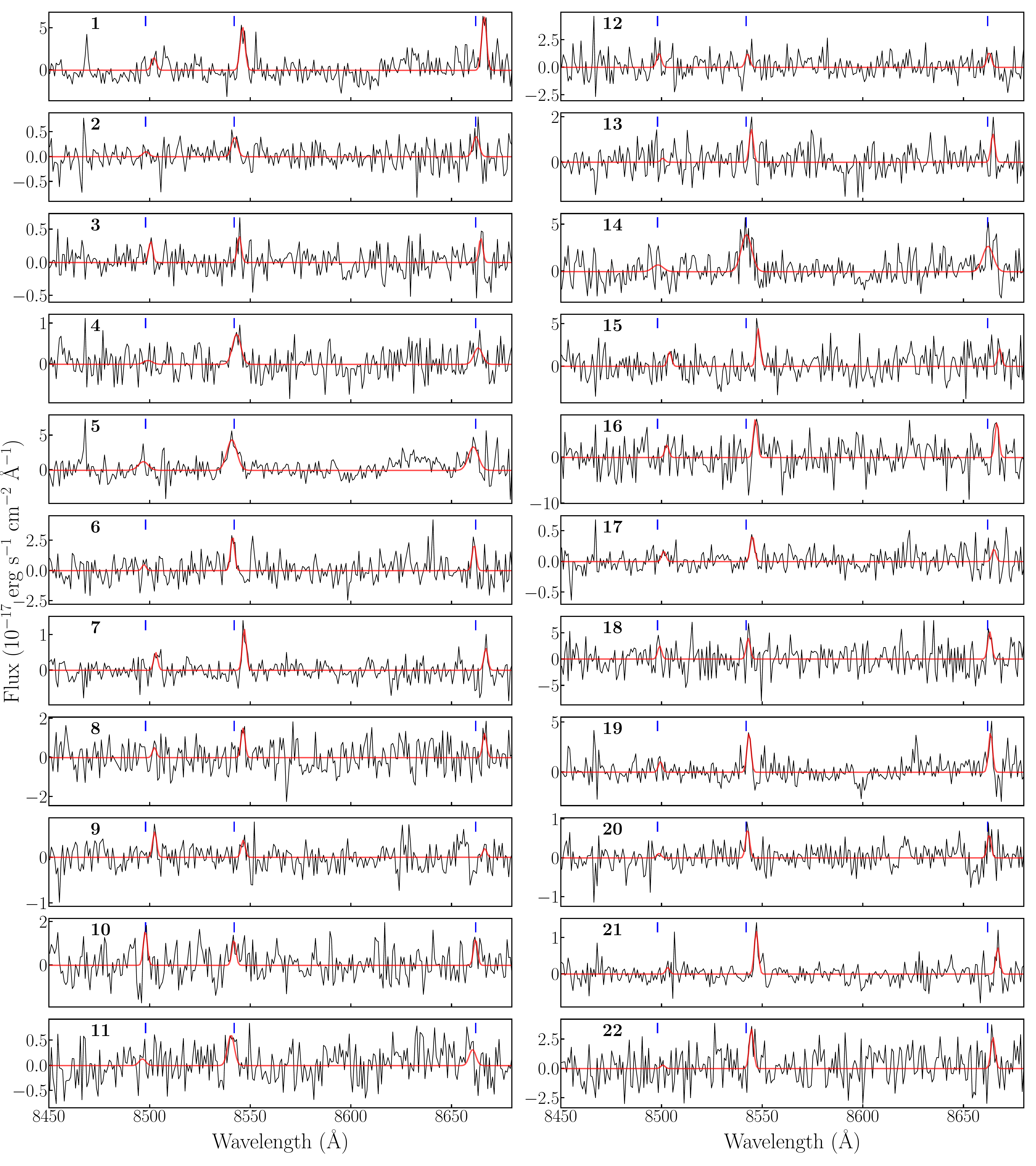}
\caption{Normalized spectra of the 22 white dwarfs with candidate gaseous debris disks in the vicinity of the \ion{Ca}{2} infrared triplet, displayed in order of their IDs (labeled at top left). For each source, the black curve shows the continuum-subtracted spectrum, the red curve shows the best-fit model comprising three Gaussians, and the vertical ticks mark the rest-frame wavelengths of the triplet.   
\label{fig-1}}
\end{figure*}

\begin{deluxetable*}{llccccccl}
\tablecaption{White Dwarfs with Candidate Gaseous Debris Disks. \label{tab-1}}
\tablehead{
 \colhead{ID} & \colhead{Name}  & \colhead{SpT} & \colhead{$T_{\rm eff}$} & \colhead{$\log g$} & \colhead{Mass}  & \colhead{Distance} & \colhead{\ion{Ca}{2}\, FWZI}  & \colhead{Comments} \\
 \colhead{} & \colhead{(DESI)}  & \colhead{} & \colhead{(K)} & \colhead{$({\rm cm\ s^{-2})}$} & \colhead{$(M_{\odot})$}  & \colhead{(pc)} & \colhead{(${\rm  km\, s^{-1}}$)} & \colhead{}
}
\startdata
1  & WD J084253.03+230025.47 & DA &  25666 & 7.77 & 0.52 & 77 & $254 \pm 30$ & DP \\
2 & WD J103812.22+825135.62 & DAZ & 28797 & 9.33 & 1.31 &  117 & $316 \pm 85$ & DP \\
3 & WD J072856.53+453123.05 & DA & 24380 & 7.73 & 0.55 &  305 & $210 \pm 57$ & DP \\
4 & WD J143645.21+154140.70 & DA & 16659 & 8.14 & 0.69 & 107 & $438 \pm 105$ & -- \\
5  & WD J151127.63+320417.92 & DA & 14770 & 8.08 & 0.66 &  34 & $522 \pm 85$ & -- \\
6  & WD J180230.44+803951.14 & DA & 25482 & 8.08 & 0.67 &  84 & $210 \pm 57$ & DP \\
7  & WD J054122.70$-$192104.74 & DB & 15266 & 7.96 & 0.59 & 132 & $210 \pm 44$ & -- \\
8& WD J084949.62+092353.46 & DA & 8569 & 7.98 & 0.58 & 97 & $210 \pm 38$ & DP, IR-ex \\
9 & WD J074709.73+162110.14 & DA & 8312 & 8.37 & 0.83 &  124 & $210 \pm 118$ & -- \\
10  & WD J091200.58+013919.51 & DA & 16554 & 8.12 & 0.68 &132 & $210 \pm 7$ & -- \\
11  & WD J093708.61+333404.69 &DA & 23135 & 6.93 & 0.26 & 312 & $351 \pm 116$ & DP \\
12 & WD J143406.77+150817.81 & DA & 14129 & 8.05 & 0.63 & 78 & $236 \pm 94$ & DP, IR-ex \\
13  & WD J001029.07+094532.54 & DA & 17604 & 7.91 & 0.56 &  111 & $210 \pm 441$ & -- \\
14 & WD J110034.24+713802.92 &DA & 39984 & 7.86 & 0.60 & 126 & $550 \pm 81$ & IR-ex \\
15  & WD J115952.05+000751.87 & DAP & 9419 & 8.65 & 1.01 &  28 & $210 \pm 104$ & -- \\
16  & WD J101606.87$-$011917.14 & DA & 7766 & 7.25 & 0.29 & 46 & $210 \pm 50$ & -- \\
17  & WD J075345.76+333527.87 & DA & 7023 & 7.74 & 0.45 & 118 & $218 \pm 55$ & -- \\
18  & WD J121129.27+572417.24 & DAZ& 5714 & 7.90 & 0.51 & 19 & $210 \pm 203$ & -- \\
19   & WD J071959.42+402122.13 & DBZ & 16953 & 7.90 & 0.56 &  59 & $210 \pm 256$ & -- \\
20  & WD J004101.22+414429.47 & DA & 8060 &  7.46 &  0.36 &  143 & $210 \pm 104$ & -- \\
21  & WD J054805.64$-$214217.63 & DA &  31292 & 8.45 & 0.92 & 193 & $210 \pm 67$ & DP \\
22 & WD J142340.72+313459.84 & DA & 27742 & 7.96 & 0.62 & 113 & $210 \pm 0$ & DP \\
\enddata
\tablecomments{Columns: (1) candidate ID; (2) DESI object name; (3) spectral type; (4) effective temperature; (5) surface gravity; (6) mass; (7) distance; (8): full-width at zero intensity of \ion{Ca}{2} $\lambda$8542 line; (9) comments: DP =  double-peaked line profile (indicative of an emitting disk), IR-ex = infrared excess detected in WISE.}
\end{deluxetable*}

Figure~\ref{fig-1} presents the spectra of the 22 white dwarfs with candidate gaseous debris disks; their properties are summarized in Table~\ref{tab-1}. We note that a significant fraction of these candidates may be false positives, and all  require follow-up observations for confirmation.

The detected \ion{Ca}{2} triplet emission lines are predominantly weak, as indicated by the best-fit model consisting of three Gaussian profiles (red lines). Those features are scarcely detectable above the noise, with signal-to-noise ratios (S/N) rarely exceeding 3. Therefore, it remains uncertain whether these are true emission lines. 
Nevertheless, the detection of three emission features with similar line widths and at a common velocity shift to their rest wavelengths, increases the likelihood that they originate from the \ion{Ca}{2} triplet. Despite this, the lines are too weak to permit dynamical modeling or detailed characterization of the putative gaseous disks.

We used the full-width at zero intensity (FWZI) to describe the line width and the maximum velocity in the disk, calculated as the $\pm3\sigma$ width of the best-fit Gaussian profile. Among the triplet, the \ion{Ca}{2} $\lambda$8542\,{\AA} line is the strongest; its FWZI values are listed in Table~\ref{tab-1}. 
More than half of the candidates exhibit FWZI values around 210 km\,s$^{-1}$, consistent with the lower limit imposed on the Gaussian standard deviation (1\,\AA).   
A small subset of systems (e.g., ID = 1), however, show asymmetric double-peaked emission profiles (labeled as `DP' in the last column of Table~\ref{tab-1}), a signature indicative of a rotating disk and similar to those observed in previously confirmed gaseous debris disk.

\begin{figure}[tbp]
\centering
\includegraphics[width=0.48\textwidth]{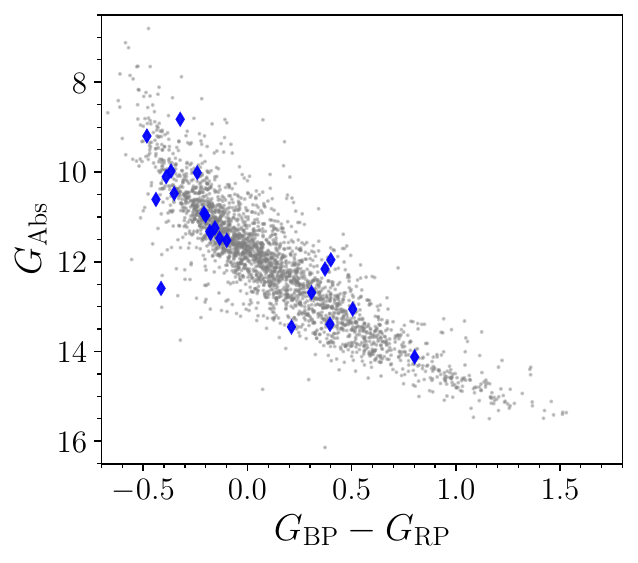}
\caption{Gaia color--magnitude diagram for the 22 candidate debris disks (blue diamonds), compared to the parent DESI EDR white dwarf sample (gray circles).
\label{fig-2}}
\end{figure}

These white dwarfs also exhibit notable spectral and temperature characteristics, though any definitive interpretation must await confirmation of the disks. For instance, most belong to the DA spectral type, which is difficult to reconcile with the presence of metallic gaseous emission. 
Additionally, the effective temperatures span a wide range ($T_{\rm eff} \sim$ 5700--40000\,K): some objects are too cool to excite or ionize calcium, while others are sufficiently hot to potentially disrupt or evaporate a gaseous disk. 
{Despite these peculiarities, Figure~\ref{fig-2} shows that the candidates are otherwise indistinguishable from the broader DESI EDR white dwarf population in the Gaia color--magnitude diagram.}

\subsection{Infrared Excess}
\begin{figure*}[htbp] 
\centering
\includegraphics[width=1\textwidth]{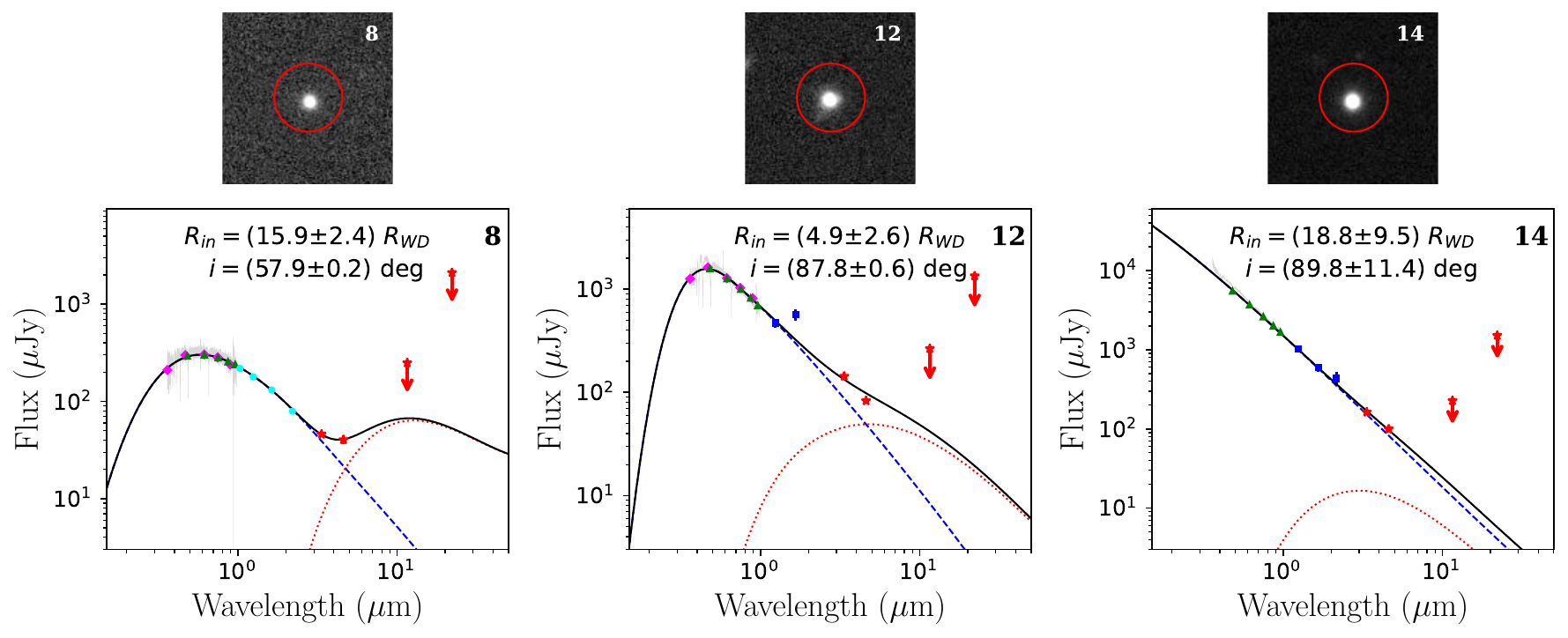}
\caption{Pan-STARRS $z$-band images (top row) and SED fits (bottom row) for the three candidate gaseous-debris-disks white dwarfs showing infrared excess. From left to right: WD~J0849+0923 (ID=8), WD~J1434+1508 (ID=12), WD~J1100+7138 (ID=14).
Each $z$-band image is centered on the DESI source position, with a red circle of 6\arcsec\,radius indicating the absence of visible contaminants within the typical WISE beam.
The SED panels show photometric data from SDSS (magenta diamonds), Pan-STARRS (green triangles), 2MASS (blue squares), UKIDSS (cyan circles), and WISE (red stars), along with the DESI spectrum in gray.
The best-fit model (black solid line) includes contributions from the white dwarf photosphere (blue dashed line) and a dusty disk (red dotted line). The derived disk parameters (inner radius $R_{\rm in}$ and inclination $i$) are listed.
\label{fig-3}}
\end{figure*}

Three of the 22 candidate gaseous disks exhibit an infrared excess ($\gtrsim 3\sigma$) in the WISE $W_1$/$W_2$ bands, specifically WD~J0849+0923, WD~J1434+1508, and WD~J1100+7138 (IDs 8, 12, and 14 in Table~\ref{tab-1}, denoted as 'IR-ex').
No significant contamination is detected within a 6\arcsec\,  aperture in the Pan-STARRS1 $z$-band images. If confirmed, these sources would represent additional cases of debris disks with gaseous and dusty components simultaneously detected. The images and SED fitting results are presented in Figure~\ref{fig-3}. 

SED modeling suggests that two of these systems are likely viewed in high inclinations (close to edge-on), although there remains a degeneracy between inclination and inner disk radius. The derived inner disk radii range from approximately $5$ to $19\,R_{\rm WD}$, corresponding to Keplerian velocities of $500-1200~{\rm km\,s^{-1}}$. 
For highly inclined disks, the high orbital velocity would result in widely separated double-peaked emission lines if the gas and dust coexist in a coplanar structure. In low S/N spectra, such broad and diluted emission features can be particularly challenging to detect, which may partly explain the weakness of the observed \ion{Ca}{2} lines, although the probability of false detections cannot be ruled out.

\section{Discussion}
\label{sec:discussion}

From a parent sample of 2706 white dwarfs, we identified 22 candidate gaseous debris disks based on the detection of the \ion{Ca}{2} emission triplet. This corresponds to a raw occurrence rate of 0.81\%, more than ten times higher than the previously reported rate of 0.067\% in the literature. 
Such a significant discrepancy suggests that the majority of these detections are likely false positives, with an expected number of true disks on the order of  $\sim 2$. 
Contamination from various sources or residual noise may account for the remaining candidates, underscoring the need for follow-up observations to confirm their nature.

\subsection{Potential Contaminants and False Positives}

The reliability of the detected \ion{Ca}{2} emission lines needs to be assessed, as several sources of contamination could lead to false positives. The following factors are considered as primary:

\begin{enumerate}[label=(\roman*), align=left]
\item{Telluric emission lines.} 
Although the DESI spectra are co-added and sky-subtracted, residual contamination from atmospheric emission remains possible, particularly for faint targets. The wavelength range of the \ion{Ca}{2} triplet coincides with ro-vibrational transitions of OH, which are challenging to subtract completely. 
A notable example is the previously reported weak \ion{Ca}{2} emission in SDSS J1344+0324 \citep[][]{2017ApJ...836...71L}, which was later attributed to imperfect sky subtraction \citep{2019AJ....158..242X}.
While most of our candidates exhibit slight redshift in their emission lines, inconsistent with rest-frame telluric features, we cannot fully rule out contamination from time-varying or imperfectly subtracted sky lines. 

\item{Binary companions.} 
Although binary systems were initially filtered from the parent sample, some may remain undetected. In such cases, 
emission lines and infrared excess could originate from stellar companions rather than circumstellar disks. 
For instance, SDSS J1144+0529, a white dwarf with weak \ion{Ca}{2} emission and infrared excess \citep{2015ApJ...810L..17G}, was later confirmed to host a low-mass companion \citep{2020MNRAS.496.5233S}.  
The significance of the \ion{Ca}{2} detections in our sample is comparable to such cases, indicating that undetected binaries may contribute to false positives. Further high-resolution imaging or spectroscopic monitoring is required to assess this possibility.
\end{enumerate}

\subsection{Comparison with the Literature}

Three of the 22 candidates in our sample have been previously discussed in the literature:

\begin{enumerate}[label=(\roman*), align=left]
\item{WD~J1802+8039 (ID=6).} 
Hubble Space telescope (HST) ultraviolet spectroscopy of this system shows no detectable silicon or carbon absorption features, which are often signposts of remnant planetary material \citep{2024ApJ...976..156O}. This is consistent with its optical spectral classification as a DA white dwarf. 
The presence of the \ion{Ca}{2} emission line in our data does not necessarily conflict with the absence of UV metal absorption, as the currently known population of gaseous debris disks is strongly biased toward metal-polluted white dwarfs.  
The processes leading to gas emission and metal absorption may originate in distinct regions or under different conditions, suggesting that detections in one wavelength regime do not guarantee signatures in the other.

\item{WD~J1100+7138 (ID=14).}  
This candidate exhibits an infrared excess in the WISE bands in our analysis. 
Earlier studies using $J/K$-band photometry reported no significant infrared excess and found no evidence of cool companions \citep{2000ApJ...540..992G}.  
Since dusty debris disks predominantly radiates at longer wavelengths, WISE provides greater sensitivity to such emission than shorter-wavelengths near-infrared surveys. 
Although previous works did not report circumstellar features \citep[e.g.,][]{1997ApJ...484..871H, 2003MNRAS.341..477B}, the WISE detection remains a robust indicator of potential dust presence.

\item{WD~J0719+4021(ID=19).}
A recent study by \citet{2025ApJ...981L...5F} reported the detection of warm circumstellar dust around this white dwarf based on mid-infrared observations with the James Webb Space Telescope (JWST).  We did not identify significant infrared excess emission in the WISE $W_1$ or $W_2$ band. 
Furthermore, the longer-wavelength $W_3$ and $W_4$ bands provide only upper limits. This indicates that the dust reported by JWST likely emits predominantly at wavelengths beyond the sensitivity range of WISE, illustrating the utility of more sensitive mid-infrared facilities in detecting cooler or weaker circumstellar material.
\end{enumerate}

The above comparison with the literature highlights the need for multi-wavelength observations to fully understand these systems and confirm weak, ambiguous features such as our emission-line candidates.

\section{Summary}
\label{sec:summary}

 We have performed a systematic search for gaseous debris disks around white dwarfs by identifying emission from the \ion{Ca}{2} infrared triplet in the DESI EDR sample. From a parent sample of 2706 white dwarfs, we identified 22 candidate systems; however, a significant fraction of them are likely false positives due to telluric residuals or unresolved binaries. 
Nevertheless, our sample includes candidates exhibiting tentative evidence of double-peaked emission line profiles and/or infrared excess, which merit follow-up observations. This work demonstrates the potential of DESI spectroscopy for the blind discovery of rare circumstellar phenomena. 
The recently released DESI DR1, which provides a substantially larger spectroscopic sample than the EDR used here, will enable better constraints on the occurrence rate and nature of gaseous debris disks.

\begin{acknowledgments}
This work is supported by the National Natural Science Foundation of China (NSFC; grant Nos. 11890692, 12133008, 12221003, and 12033004), the China Manned Space Program (grant No. CMS-CSST-2025-A10), the National Key R\&D Program of China (grant No. 2023YFA1607904),
the Fundamental Research Fund for the Central Universities of China (Grant No. 20720230016),  Fujian Provincial Natural Science Foundation of China (Grant No. 2024J08001), and the Natural Science Foundation of Xiamen, China (No. 3502Z202472007). 
J.G. is supported by the NSFC under grants No. 12203006 and the Young Scholar Program of Beijing Academy of Science and Technology (No. 25CE-YS-02).
X.J. acknowledges the support from the grant No. JAT241087 provided by the Fujian Provincial Department of Education.
\end{acknowledgments}

\software{Astropy \citep{2018AJ....156..123A}, 
Matplotlib \citep{2007CSE.....9...90H}, 
SciPy \citep{2020NatMe..17..261V}.
          }

\bibliography{ms}{}
\bibliographystyle{aasjournal}

\end{CJK*}
\end{document}